\documentclass[12pt]{iopart}

\usepackage{mathrsfs}
\usepackage{graphicx}% Include figure files
\usepackage{subfigure}
\usepackage{amsfonts}
\usepackage{amssymb}
\begin{document}

\title[]{Integrable nonlinear Klein-Gordon systems with $\mathcal{PT}$ nonlocality and/or space-time exchange nonlocality}

\author{Man Jia \& S. Y. Lou}

\address{School of Physical Science and Technology, Ningbo University, Ningbo 315211, P. R. China}
\ead{jiaman@nbu.edu.cn}
\vspace{10pt}
%\begin{indented}
%\item[]August 2017
%\end{indented}

\begin{abstract}
In additional to the parity ($\mathcal{P}$) symmetric, time reversal ($\mathcal{T}$) symmetric, and $\mathcal{PT}$ symmetric nonlocal integrable systems, some other types of nonlocal integrable Klein-Gordon models with the space-time exchange nonlocality and the moving nolocality are proposed. The Lax pairs of the established nonlinear nonlocal Klein-Gordon equations are explicitly given. A special soliton solution, composed of $\mathcal{PT}$-symmetric part and $\mathcal{PT}$-antisymmetric part, is illustrated with the shape change.
\end{abstract}

%
% Uncomment for keywords
%\vspace{2pc}
%\noindent{\it Keywords}: Travelling wave solution, few-cycle-pulse solitons, soliton molecules;
%
% Uncomment for Submitted to journal title message
%\submitto{\JPA}
%
% Uncomment if a separate title page is required
%\maketitle
%
% For two-column output uncomment the next line and choose [10pt] rather than [12pt] in the \documentclass declaration
%\ioptwocol
%

Since the first nonlocal nonlinear Sch\"{o}dinger (NNLS) system
\begin{eqnarray}
i A_t=A_{xx} \pm 2 A^2 B, \qquad B= \hat{P} \hat{C}A=A^{\ast}(-x,t), \label{nnls}
\end{eqnarray}
was introduced by Ablowitz and Musslimani in \cite{PhysRevLett.110.064105}, the research on the nonlocal nonlinear systems has been a hot spot for serval years. Many novel nonlocal nonlinear systems, such as the nonlocal Korteweg-de Vries equation (KdV) \cite{louhuang2017}, \cite{doi:10.1063/1.5051989}, \cite{JIA20181157}, \cite{Lou_2020}, the nonlocal modified Korteweg-de Vries equation (mKdV) \cite{JI2017699}, \cite{liABmkdv}, \cite{tang2018}, \cite{JI2017973}, \cite{jia2020}, the other types of NNLS systems \cite{RevModPhys.88.035002}, \cite{SONG201713}, \cite{Fokas_2016}, \cite{PhysRevE.98.042202}, \cite{doi:10.1063/1.4960818}, \cite{https://doi.org/10.1111/sapm.12153}, \cite{PhysRevE.91.023201}, the discrete nonlocal NLS systems \cite{PhysRevE.90.032912}, the nonlocal Davey-Stewartson systems \cite{doi:10.1063/1.4817345}, the nonlocal peakon systems \cite{lou100201}, the nonlocal Boussinesq-KdV type systems \cite{https://doi.org/10.1111/sapm.12265}, the nonlocal Sawada-Kotera system \cite{zhaoabsk2020}, the nonlocal Sine-Gordon/Sinh-Gordon equation with nonzero boundary conditions \cite{https://doi.org/10.1111/sapm.12222} and so on, have been established by using different methods. The established nonlocal nonlinear systems are proved to be integrable with the Lax pair, infinitely many conservation laws and generally can be solved by the inverse scattering method \cite{PhysRevLett.110.064105}, \cite{doi:10.1063/1.5051989}, \cite{PhysRevE.98.042202}, \cite{Ablowitz_2016}, Darboux transformation  \cite{PhysRevE.91.033202}, the bilinear method \cite{https://doi.org/10.1111/sapm.12265}, the symmetries and nonlocal symmetries \cite{JIA20181157}, \cite{Lou_2020}, \cite{jia2020}, etc. A wide array of novel nonlinear phenomena and new properties, including the prohibitions caused by nonlocality \cite{https://doi.org/10.1111/sapm.12265}, the shape of the solitons dependent on the spectral parameter \cite{liABmkdv}, the soliton transformed from dark to bright during propagation \cite{jia2020}, the special singularities of the solitons at some sites \cite{JI2017973}, the distinctive soliton dynamics \cite{RevModPhys.88.035002}, the symmetry breaking of solitons \cite{PhysRevE.91.023201}, has been revealed for the special nonlocalities in optics, Bose-Einstein condensates, atmosphere and fluid, electronic circuits, and many other branches of physical fields.

The operators $\hat{P}$ and $\hat{C}$ in eq.~(\ref{nnls}) come out from quantum mechanics are referred to the parity and charge conjugation (complex conjugate in mathematics), respectively. The parity operator produces a space reflection on the particle and antiparticle states of the field. The charge conjugation operator transforms a particle into its corresponding antiparticle. The NNLS equation eq.~(\ref{nnls}) is believed to possess the $\mathcal{PC}$-symmetry \cite{doi:10.1063/1.5051989}, \cite{Lou_2020}, but was previously considered to be $\mathcal{PT}$ ($\hat{P}\hat{T}$, meaning parity-time reversal) symmetric. Because most of the physical systems possess combined $\mathcal{CPT}$ symmetry (parity and/or time-reversal and/or charge-conjugate) \cite{de_Almeida_2014}, \cite{PhysRevLett.89.231602}, \cite{PhysRevD.104.023509}, \cite{PhysRevD.103.106015}, \cite{doi:10.1142/S0219887810004816}, the applications of $\mathcal{PT}$, $\mathcal{PC}$ symmetries and their suitable combinations to other branches of physics and mathematics then quickly followed.

These developments motivated further extension by using the concept of symmetries because symmetries are not only the basic principles to establish physical models but also the key points to solve complicated scientific problems. Among all the foundational symmetries of the physical systems, the exchange symmetry of space-time ($\hat{E}_{x,\ t}\ \{x,\ t\}=\{t,\ x\}$, the exchange of $x$ and $t$) is extraordinary interesting and important due to the wide applications in the special and general relativity, string theory, particle physics, classical and quantum fields \cite{LUCIOM1996150,PhysRevD.104.104046,doi:10.1063/5.0039991,PhysRevA.73.053411,doi:10.1063/1.4961431}.

It is known that the $(1+1)$-dimensional nonlinear Klein-Gordon (KG) equation in light cone coordinate \begin{eqnarray}
u_{xt}=F(u)\label{kg}
\end{eqnarray}
is integrable only for the Liouville case ($F(u)=\exp(u)$), the Sine-Gordon (sG) case ($F(u)=\sin(u)$), the Sinh-Gordon case ($F(u)=\sinh(u)$) and the Tzitzeca case ($F(u)=\exp(u)+\exp(-\frac12u)$).
The real nonlinear KG equation (\ref{kg}) is $\mathcal{PT}$ invariant and relativistically invariant, i.e., invariant under the transformations $\{x,\ t\} \rightarrow \{-x,\ -t\}$ and $\{x,\ t\} \rightarrow \{t,\ x\}$.

%The positive/negative hierarchy duality is studied with the help of the relativistic invariant symmetry. The Lorentz symmetry in ghost-free massive gravity investigated in \cite{PhysRevD.104.104046} provides a guide to choices of parameters in the potential for massive gravity. The Lorentz invariant Lagrangian is constructed to include the quantum nonlinearity parameter dependency of the invariant photon mass in \cite{PhysRevD.104.104046}. A multi-time formalism and a Lorentz-invariant rule for the coordination of the spacetime points is employed on the individual particle trajectories \cite{Horton_2001}.

Due to the significance of relativistically invariant or space-time exchange invariant, more researches have been done to explore the relativistically invariant physical systems and to find new phenomena related to the space-time exchange invariant in different branches of physics\cite{lou2021JHEP}. But to the best of our knowledge, there are few results about nonlocal nonlinear systems with space-time exchange nonlocality. Motivated by the extensions of
$\mathcal{PT}$ symmetric nonlocal models from optical spatial soliton systems, both the $\mathcal{PT}$ symmetry and the space-time exchange symmetry will be used to construct some new nonlocal integrable KG systems in this manuscript.

To construct the nonlocal integrable systems with space-time exchange nonlocality, we start from a special integrable nonlinear Klein-Gordon (KG) equation \cite{Fordy1980}
\begin{eqnarray}
u_{xt}=\textrm{e}^{2u}+\textrm{e}^{-u}\cos(3v), \qquad v_{xt}=-\textrm{e}^{-u}\sin (3v),\label{sguv}
\end{eqnarray}
which is closely related to the periodic Toda lattice equations with $u \equiv u(x,\ t)$ and $v \equiv v(x,\ t)$. The nonlinear KG equation eq.~(\ref{sguv}) stems from the special structure of the $3 \times 3$ following linear spectral problem
\begin{eqnarray}
\varphi_x=M\varphi, \qquad \varphi_t=N \varphi,\label{lax}
\end{eqnarray}
where $\varphi=(\varphi_1,\ \varphi_2, \ \varphi_3)^T$ and
\begin{eqnarray}
&& M=\left(\begin{array}{ccc}
    u_x+iv_x & \lambda & 0 \\
    0 & -2iv_x & \lambda \\
    \lambda & 0 & -u_x+iv_x
  \end{array}
\right), \nonumber \\ &&
N=\lambda^{-1}\left(\begin{array}{ccc}
          0 & 0 & \textrm{e}^{2u} \\
          -\textrm{e}^{-u-3iv}& 0 & 0 \\
          0 & -\textrm{e}^{3iv-u} & 0
        \end{array}
\right),\label{mn}
\end{eqnarray}
with the arbitrary spectral parameter $\lambda$. The exactly single kink solution, the momentum integral and the superposition formula of the nonlinear KG equation eq.~(\ref{sguv}) are all given in \cite{Fordy1980}.

It is clear the nonlinear KG equation eq.~(\ref{sguv}) possesses a discrete symmetry group that can be expressed by
\begin{eqnarray}
\mathcal{G} \equiv \{1, \ \hat{P}\hat{T}, \ \hat{E}_{x,t},\ \hat{P}\hat{T}\hat{E}_{x,t} \},\label{g}
\end{eqnarray}
where the operators $\hat{P}$, $\hat{T}$ and $\hat{E}_{x,t}$ are the parity, time reversal and the exchange of $x$ and $t$ defined by
\begin{eqnarray}
\hat{P}x=-x, \qquad \hat{T}t=-t, \qquad \hat{E}_{x,t}\{x,\ t\}=\{t,\ x\},
\end{eqnarray}
respectively. The discrete symmetry group eq.~(\ref{g}) denotes the nonlinear KG equation eq.~(\ref{sguv}) is invariant under the transformations $x \rightarrow -x,\ t \rightarrow -t$ and/or $x \rightarrow t,\ t \rightarrow x$ and/or $x \rightarrow -t,\ t \rightarrow -x$. In other words, the nonlinear KG equation eq.~(\ref{sguv}) possesses $\mathcal{PT}$-symmetry and/or Lorentz symmetry (or, relativistically invariance).

Introducing
\begin{eqnarray}
u=p+q, \qquad v=p-q, \qquad q=\hat{g} p, \qquad \hat{g} \in \mathcal{G},
\end{eqnarray}
to the normal nonlinear KG equation eq.~(\ref{sguv}), a particular nonlocal nonlinear KG system is directly constructed
\begin{eqnarray}
p_{xt}=\frac{1}{2}\left[\cos(3p-3q)-\sin(3p-3q)\right]\textrm{e}^{-(p+q)}+ \frac{1}{2}\textrm{e}^{2(p+q)},\label{nkg1}
\end{eqnarray}
with $q= \hat{g} p$, $\hat{g} \in \mathcal{G}$.

Because $\hat{g}$ can be take any one element of the discrete symmetry group eq.~(\ref{g}), nonlinear KG equation eq.~(\ref{nkg1}) includes four different cases:\\
\em Case 1. \rm The local Tzitzeca equation,
\begin{eqnarray}
u_{xt}=\textrm{e}^{-u}+ \textrm{e}^{2u},\ q=p=\frac{u}2,\label{nkg11}
\end{eqnarray}
can read off from (\ref{nkg1}) by taking $\hat{g}=1$. \\
\em Case 2. \rm The integrable KG equation
\begin{eqnarray}
&& p_{xt}=\frac{1}{2}\left\{\cos[3p(x,t)-3p(-x,-t)]-\sin[3p(x,t)-3p(-x,-t)]\right\}\nonumber \\ && \qquad \times
\textrm{e}^{-[p(x,t)+p(-x,-t)]}+ \frac{1}{2}\textrm{e}^{2[p(x,t)+p(-x,-t)]},\label{nkg12}
\end{eqnarray}
with $\mathcal{PT}$ nonlocality is related to (\ref{nkg1}) by fixing $\hat{g}=\hat{P}\hat{T}$. \\
\em Case 3. \rm If $\hat{g}$ is selected as the space-time exchange operator $\hat{E}_{x,t}$, then (\ref{nkg1}) becomes an integrable KG equation with the space-time exchange nonlocality,
\begin{eqnarray}
&& p_{xt}=\frac{1}{2}\left\{\cos[3p(x,t)-3p(t,x)]-\sin[3p(x,t)-3p(t,x)]\right\}\nonumber \\ && \qquad \times \textrm{e}^{-[p(x,t)+p(t,x)]}
+ \frac{1}{2}\textrm{e}^{2[p(x,t)+p(t,x)]}.\label{nkg13}
\end{eqnarray}
\em Case 4. \rm If $\hat{g}=\hat{E}_{x,t}\hat{P}\hat{T}$ is selected, then (\ref{nkg1}) becomes an integrable KG equation with the space-time exchange nonlocality companied by the $\mathcal{PT}$ nonlocality,
\begin{eqnarray}
&& p_{xt}=\frac{1}{2}\left\{\cos[3p(x,t)-3p(-t,-x)]-\sin[3p(x,t)-3p(-t,-x)]\right\}\nonumber \\ && \qquad \times\textrm{e}^{-[p(x,t)
+p(-t,-x)]}+ \frac{1}{2}\textrm{e}^{2[p(x,t)+p(-t,-x)]}.\label{nkg14}
\end{eqnarray}

It is clear that the nonlocal KG equation eq.~(\ref{nkg1}) is Lax integrable because it can be derived from the $3\times 3$ compatibility condition
\begin{eqnarray}
\partial_x \varphi =M_1 \varphi,\qquad
\partial_t \varphi=N_1 \varphi,
\end{eqnarray}
where
\begin{eqnarray}
&& M_1=\left(\begin{array}{ccc}
                  (p+q)_x+i (p-q)_x & \lambda & 0 \\
                  0 & -2 i(p-q)_x  & \lambda \\
                  \lambda & 0 & -(p+q)_x+i (p-q)_x \\
                \end{array}
\right),\nonumber \\ && N_1=\frac{1}{\lambda}\left(\begin{array}{ccc}
                  0 & 0 & \textrm{e}^{2(p+q)} \\
                  -\textrm{e}^{-(p+q)-3i (p-q)} & 0  & 0 \\
                  0 & -\textrm{e}^{-(p+q)+3i (p-q)} & 0 \\
                \end{array}
\right).
\end{eqnarray}

In fact, starting from coupled integrable nonlinear KG equations one can derived various other nonlocal KG equations with different nonlocalities such as the $\mathcal{PT}$ nonlocality, and/or the space-time exchange nonlocality. Here we just list some other types of integrable nonlinear KG equations without details on the derivations because of the similarities.

A). A variant form of (\ref{nkg1}) reads
\begin{eqnarray}
&& p_{xt}=\frac{1}{2}\left[\sin(3p-3q)-\cos(3p-3q)\right]\textrm{e}^{-(p+q)}+ \frac{1}{2}\textrm{e}^{2(p+q)}, \nonumber \\ && q=\hat{g} p, \qquad \hat{g} \in \mathcal{G}, \label{nkg2}
\end{eqnarray}
which relates to the following linear spectral problem:
\begin{eqnarray}
&& \varphi_x =M_2 \varphi,\qquad \varphi_t =N_2\varphi,\nonumber \\ && M_2=\left(\begin{array}{ccc}
                  (p+q)_x+i (p-q)_x & \lambda & 0 \\
                  0 & -2 i(p-q)_x  & \lambda \\
                  \lambda & 0 & -(p+q)_x+i (p-q)_x \\
                \end{array}
\right),\nonumber \\ && N_2=\frac{1}{\lambda}\left(\begin{array}{ccc}
                  0 & 0 & \textrm{e}^{2(p+q)} \\
                  \textrm{e}^{-(p+q)-3i (p-q)} & 0  & 0 \\
                  0 & \textrm{e}^{-(p+q)+3i (p-q)} & 0 \\
                \end{array}
\right).
\end{eqnarray}

B). The second type of nonlocal integrable KG equation reads
\begin{eqnarray}
&& p_{xt}=\frac{1}{2}\left[\sinh(3p-3q)-\cosh(3p-3q)\right]\textrm{e}^{-(p+q)}+ \frac{1}{2}\textrm{e}^{2(p+q)}, \nonumber \\ && q=\hat{g} p, \qquad \hat{g} \in \mathcal{G}, \label{nkg3}
\end{eqnarray}
which comes from the compatibility condition by $\varphi_{xt}=\varphi_{tx}$, and $\lambda$ is the spectral parameter with
\begin{eqnarray}
&& \varphi_x=\left(\begin{array}{ccc}
                  2p_x & \lambda & 0 \\
                  0 & -2 (p-q)_x  & \lambda \\
                  \lambda & 0 & -2q_x \\
                \end{array}
\right)\varphi,\nonumber \\ && \varphi_t =\frac{1}{\lambda}\left(\begin{array}{ccc}
                  0 & 0 & \textrm{e}^{2(p+q)} \\
                  \textrm{e}^{-4p+2q} & 0  & 0 \\
                  0 & \textrm{e}^{2p-4q }& 0 \\
                \end{array}
\right)\varphi
\end{eqnarray}

C). The third type of nonlocal KG equation has the form
\begin{eqnarray}
&& p_{xt}=\textrm{e}^{2p}+\textrm{e}^{-(p+q)}, \qquad q=\hat{g} p, \qquad \hat{g} \in \mathcal{G}, \label{nkg4}
\end{eqnarray}
which is integrable guaranteed by the following Lax pair with $\varphi= (\varphi_1,\ \varphi_2,\ \varphi_3,\ \varphi_4)^T$,
\begin{eqnarray}
&& \partial_x \varphi =M_4 \varphi,\qquad
\partial_t \varphi =N_4 \varphi,\nonumber \\ && M_4=\left(
                                                            \begin{array}{cccc}
                                                              -p_x & \lambda & 0 & 0 \\
                                                              0 & p_x & \lambda & 0 \\
                                                              0 & 0 & -q_x & \lambda \\
                                                              \lambda & 0 & 0 & q_x \\
                                                            \end{array}
                                                          \right)
,\nonumber \\ && N_4=\frac{1}{\lambda} \left(
                                         \begin{array}{cccc}
                                           0 & 0 & 0 & \textrm{e}^{-(p+q)} \\
                                           \textrm{e}^{2p} & 0 & 0 & 0 \\
                                           0 & \textrm{e}^{-(p+q)} & 0 & 0 \\
                                           0 & 0 & \textrm{e}^{2q} & 0 \\
                                         \end{array}
                                       \right)
.
\end{eqnarray}

D). The last type of nonlocal KG equation with $\mathcal{PT}$ nonlocality and/or the space-time exchange nonlocality has the form
\begin{eqnarray}
&& p_{xt}=-\frac{\cot (p+q) \csc^2 (p+q) }{2}\left[(p-q)_x (p-q)_t\right] - \frac{m}{16} \sin(2p+2q)\nonumber \\ && \qquad +2 (p_x p_t-q_x q_t)\textrm{csc} (2p+2q), \nonumber \\ && q=\hat{g} p, \qquad \hat{g} \in \mathcal{G}, \label{nkg5}
\end{eqnarray}
with the original integrable coupled local system presented in \cite{PhysRevD.14.1524} describing the classical theory of one-dimensional extended objects interacting through the scalar field.

To end this short paper, we try to search for the exact solutions to the nonlocal integrable KG systems. Here we just take the nonlocal KG equation eq.~(\ref{nkg1}) with $q=p(-x,\ -t)$ as a special example.

\begin{figure}
\begin{center}
\subfigure{
\includegraphics[width=0.35\textwidth]{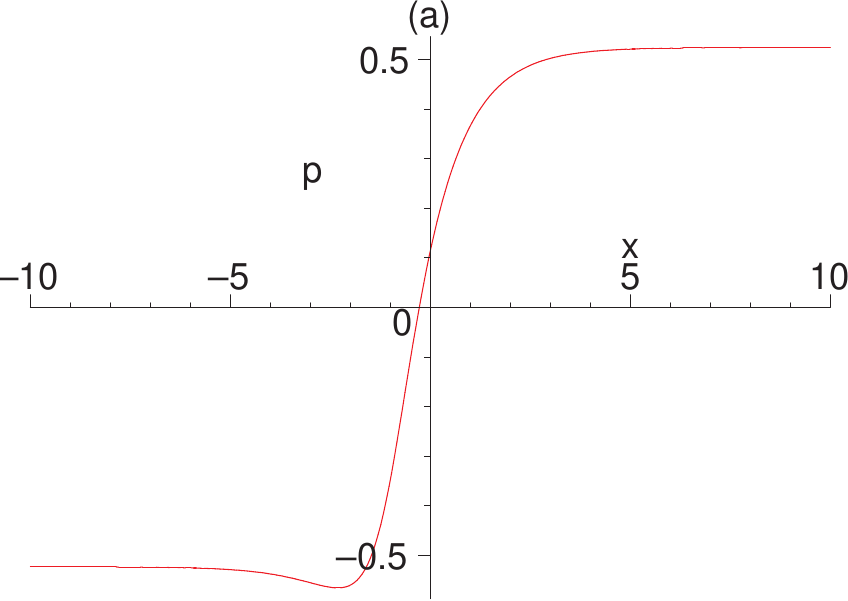}
\includegraphics[width=0.35\textwidth]{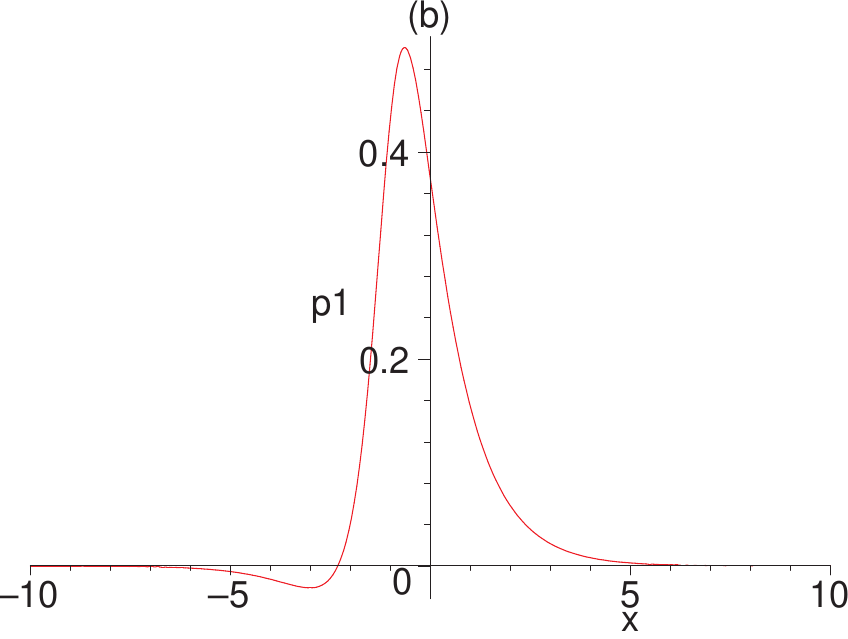}}
\caption{The exact solution to the nonlocal integrable KG system eq.~(\ref{nkg1}) with $k=1$ and $\eta_0=1$ given by eq.~(\ref{solnkg1}) for (a) the solution and (b) the potential $p_1=p_x$. \label{fig1}}
\end{center}
\end{figure}

Because $\hat{g}^2=1$, any function can be expanded as a summation of a symmetric part and an antisymmetric part with respect to $\hat{g}$ \cite{doi:10.1063/1.5051989}. Thus,
the solution to the nonlocal KG equation eq.~(\ref{nkg1}) can be written as
\begin{eqnarray}
p=A+B,\label{p1}
\end{eqnarray}
where $A$ is the symmetric part and $B$ is the anti-symmetric part with respect to $\hat{g}=\hat{P}\hat{T}$, i.e.,
\begin{eqnarray}
 \hat{P}\hat{T}A=A, \quad  \hat{P}\hat{T}B=-B.
\end{eqnarray}
Thus, the solution of the $q$-part now is
\begin{eqnarray}
q=\hat{P}\hat{T} p=A-B.\label{q1}
\end{eqnarray}
Because the exact solution to the normal KG equation eq.~(\ref{sguv}) has been presented \cite{Fordy1980}, by selecting suitable $A$ and $B$, the exact solution to the nonlocal integrable KG system eq.~(\ref{nkg1}) can be derived directly. A special selection for the single soliton of $A$ and $B$ is
\begin{eqnarray}
&& A=\frac{1}{4}\ln\left[\frac{\cosh\left(\frac{3\eta}{2}\right)}{4\cosh^3\left(\frac{\eta}{2}\right)}\right],\qquad B=\frac{1}{2}\arctan \left[\sqrt{3} \tanh \left(\frac{\eta}{2}\right)\right], \nonumber \\ && \eta=k x+3k^{-1}t,\label{solnkg1}
\end{eqnarray}
with $k$ being an arbitrary constant. The structure of the solution eqs.~(\ref{p1}) and (\ref{solnkg1}) and the corresponding potential $p_x$ is illustrated in fig.~\ref{fig1}. It can be seen that the kink solution is slightly deformed compared to the normal kink and the soliton is asymmetric due to the nonlocality caused by the parity and time reversal.

It is also interesting that if we fix $k=\sqrt{3}$ in (\ref{solnkg1}), the solution (\ref{p1}) with (\ref{solnkg1}) is also a single soliton like solution of the nonlocal KG equation (\ref{nkg14}) with the space-time exchange nonlocality companied by the parity and time reversal nonlocality.

In summary, our results shows two perspectives of considerable practical interest. One is the establishments of the novel nonlinear integrable KG systems related to the generalised Toda lattices that possess the $\mathcal{PT}$ nonlocality and/or the space-time exchange nonlocality. Though the $\mathcal{CPT}$ symmetries and their combinations have been used to construct nonlocal physical systems, this is the first time that the space-time exchange symmetry are theoretically introduced to the nonlinear integrable systems. The possible applications, the properties and the potential new phenomena of the space-time symmetry need further progresses on both theories and experiments.

The other is the emergence of the abundant integrable models due to the introduces of the nonlocalities. Because the discrete group $\mathcal{G}$ has four elements, the identity, the parity-time reversal $\hat{P}\hat{T}$, the exchange of space-time $\hat{E}_{x,t}$, the parity-time reversal and exchange of space-time $\hat{P}\hat{T}\hat{E}_{x,t}$, each of the nonlocal models eqs.~(\ref{nkg1}), (\ref{nkg2}) (\ref{nkg3}), (\ref{nkg4}), (\ref{nkg5}) includes four various types of integrable systems related to the KG equations. More research, such as the multisoliton solutions, the superposition formula of the integrable models should be done in future work.
\section*{Acknowledgement}
The authors acknowledge the support of NNSFC (No. 11975131) and K. C. Wong Magna Fund in Ningbo University.

%\bibliography{ref}
%\bibliographystyle{iopart-num}
\providecommand{\newblock}{}

\end{document}